\documentclass[]{revtex4}

\usepackage[english]{babel}  
\usepackage{amssymb,amsmath,amsthm}
\usepackage{enumerate}
\usepackage{array}
\usepackage{graphicx}
\usepackage{makeidx}  
\makeindex 

\newtheorem{theorem}{Theorem}
\newtheorem{lemma}{Lemma}

\newcommand{\CC}{{\mathbb C}}

\newcommand{\RR}{{\mathbb R}}

\newcommand{\pone}{\tilde{\dot P}}
\newcommand{\ptwo}{\tilde{\dot{\pone}}}

\newcommand{\Om}{\Omega}
\newcommand{\Hi}{{\cal H}}
\newcommand{\Ra}{{\cal R}}
\newcommand{\dist}{{\rm dist}}
\newcommand{\tp}{\thinspace}
\newcommand{\ra}{\rangle}

\newcommand{\herm}[1]{{#1}^\dagger}	
\newcommand{\ub}[1]{ {#1} \Bigr|_\mathrm{u.b.}}	

\begin{document}

%

\title{Bounds for the adiabatic approximation with applications to quantum computation}

\author{Sabine Jansen}
\email{jansen@math.tu-berlin.de}
\affiliation{Institut f\"ur Mathematik, MA~7-2,
	  TU Berlin, Stra\ss e des 17.~Juni~136, D-10623 Berlin, Germany}
\author{Mary-Beth Ruskai}
\email{Marybeth.Ruskai@tufts.edu}
\affiliation{Department of Mathematics,  Tufts University, Medford, MA 02155, USA\\
	\medskip
	Dedicated to Barry Simon on the occasion of his 60th birthday	
}
\author{Ruedi Seiler}
\email{seiler@math.tu-berlin.de}
\affiliation{Institut f\"ur Mathematik, MA~7-2,
	  TU Berlin, Stra\ss e des 17.~Juni~136, D-10623 Berlin, Germany}

\begin{abstract}
	We present straightforward proofs of estimates used in the adiabatic
	approximation. The gap dependence is analyzed explicitly.
	We apply the result to interpolating Hamiltonians of
	interest in quantum computing.
\end{abstract}

\pacs{03.65.Db, 03.67.Lx}

\maketitle


\section{Introduction}

The quantum adiabatic approximation has a long history, going back to Born and
Fock \cite{bf} early in the development of quantum theory.   Recently, the
realization that the adiabatic approximation could be used as the basis for a
method of quantum computation \cite{fgg1} has generated a resurgence of
interest in this topic. 

Despite the existence of an extensive literature \cite{asy1,hj,kato,n,jp} on
rigorous proofs of estimates needed to justify the adiabatic approximation,
doubts have been raised about its validity \cite{ms} leading to confusion about
the precise conditions needed to use it \cite{ar,tkso,mep}.   In part, this is
because some papers emphasize different aspects, such as the asymptotic
expansion, the replacement of the requirement of a nondegenerate ground state
by a spectral projection separated from the rest of the spectrum, dependence of
first order estimates on the spectral gap, and even extensions to systems
without a gap \cite{ae}.  Moreover, the qualitative gap conditions frequently presented
in elementary texts \cite{lanlif,m} are known to be insufficient.
 
In this paper, we present a straightforward, yet rigorous, proof of the
asymptotic estimates in a form which makes explicit the relevance of the gap.
The proof given here is based on the one in Ref.~\onlinecite{asy1} with the modifications
introduced in Refs.~\onlinecite{asy2} and~\onlinecite{ks}.  The main idea is to consider the
physical time evolution as a perturbation of the adiabatic time evolution.
This leads naturally to   an integration by parts formula which is the main
technical tool.
  
Roughly speaking, the adiabatic approximation says that when a Hamiltonian
changes slowly in time, the corresponding time evolution  approximately
preserves spectral subspaces.   In particular an eigenstate   $\psi(0)$ evolves
with high probability to the eigenstate $\psi(t)$ of the instantaneous
Hamiltonian $H(t)$ if the energy curve $E(t)$ does not come too close to any
other energy level of $H(t)$.  To make this precise, one considers the family
of Hamiltonians $H(t/\tau)$, where $t$ should be thought of as the
``external'' clock time and the parameter $\tau$ defines a time scale which
allows tuning of the speed of change of the Hamiltonian. It gives rise to a
scaled time  $s =t/\tau$.  Using this notation time evolution is
defined in terms of the family of time dependent Schr\"{o}dinger equations  
\begin{equation}\label{unscaled}
	i\partial _t \phi_\tau (t)   =   H(t/\tau)\phi_\tau (t) 
\end{equation}
or equivalently
\begin{equation} \label{schroed}
 	i\partial _s\psi_\tau (s)  =  \tau H(s)\psi_\tau (s).
\end{equation}
In the most common scenario, the adiabatic theorem is an asymptotic expansion
in $1/\tau$ for the error involved in estimating the time-evolved ground
state of $H(0)$ by the ground state of $H(s)$.\\

{\emph{Remark: Allowed time dependence.}} The adiabatic theorems give results
when the time dependence of the Hamiltonian in the unscaled time $t$ is of the
form $H(t/\tau)$ for some $\tau$-independent family of operators $H(\cdot)$.

It is important to understand that this excludes
Hamiltonians with several independent time scales. In particular, the example
by Marzlin and Sanders \cite{ms} which involves a Hamiltonian $H(t/\tau,
\omega_0 t)$ with fixed frequency $\omega_0$ and big time scale $\tau$ does
\emph{not} belong to this class. To see this,
notice that in terms of the dimensionless time $s=t/\tau$, the Hamiltonian is
$H(s, \omega_0 \tau s)$ which is $\tau$-dependent unless $\omega_0\tau$ is
taken to be constant. \\

In adiabatic quantum computation \cite{fgg1,fgg2}, the Hamiltonians
of interest are usually interpolating Hamiltonians of the form
  \begin{equation}
      H(s) = [1-f(s)] H_0  + f(s)H_1.
   \end{equation}
Generalizations of the form
  \begin{equation}
      H(s) = [1-f(s)] H_0  + f(s)H_1 +  k(s) H_2
   \end{equation}
with $k(0) = k(1) = 0$ have also been considered \cite{fgg3}.
In most applications of the adiabatic theorem to quantum computation, there is
another parameter, $n$, which describes the size of the system, and one is
interested in the behavior of a family of quantum systems when $n$ is large. In
particular, one  is often interested how certain quantities, such as the
running time of a computer program, grow (or decrease)
with~$n$.  This brings   a new element
into the discussion of the adiabatic approximation and requires a careful
analysis of the error bound in terms of the gap of a family of quantum systems.
We present such estimates in Theorems~\ref{theorem-gap1} and~\ref{theorem-gap2}
of this paper. 

The essential assumption in the adiabatic theorem as it is presented here 
is that the spectrum $\sigma(H(s))$ has
a band associated with the spectral projection $P(s)$, which is
separated by a gap $g(s)$ from the rest.

\begin{figure}
 	\begin{center}
 	 	\includegraphics{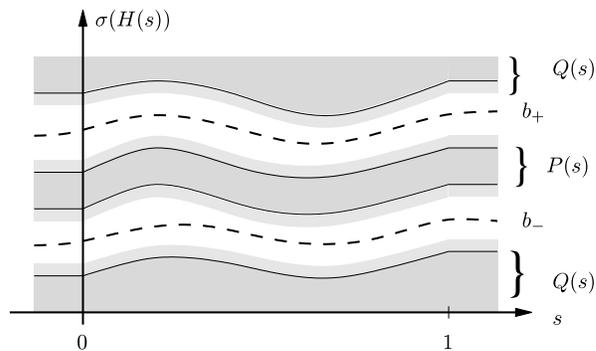}
 	\end{center}
	\caption{\small
	Energy band bordered by gaps. Shaded areas correspond to 
 	spectrum of $H(s)$. The projection $P(s)$ is associated with 
 	$\sigma(H(s))\cap [b_-(s),b_+(s)]$, and $Q(s)=1-P(s)$.
	}
\end{figure}

This setting leads to an adiabatic approximation where the error terms are
$O(1/\tau^q)$ for some $q \geq 1$. There is a weaker form of the adiabatic
theorem, where one does not require a spectral gap due to Avron and Elgart \cite{ae} and
Bornemann \cite{bornemann}. In this case the estimate on the error term is
$o(1)$ as $\tau$ goes to infinity. 

There are several different viewpoints for a discussion of the quantum
adiabatic theorem. Each one offers interesting insight.  We shall only mention
them  briefly, since the main objective of this note is to give a short and
concise proof of the adiabatic theorem in a setting which seems natural.

Berry \cite{berry} pointed out that time evolution of a quantum system in the
adiabatic limit becomes geometric. Simon \cite{simon} noted that Berry's
discovery can be interpreted as a parallel transport in a vector bundle 
with Berry's phase as  the corresponding holonomy.  Kato \cite{kato}  used
the adiabatic time evolution as a powerful tool in his perturbation theory of
linear operators.  Born and Oppenheimer \cite{bo} used the adiabatic
approximation in order to separate fast and slow motion in molecules thereby
explaining the qualitative picture of their spectra.  Niu and Thouless
\cite{tn} and Avron and Seiler \cite{asy1} showed that conductance in quantum
Hall systems is defined in the adiabatic limit and is a topological number. 

\section{Statement of results}

In the following, we will use the following notational conventions: The letter
$P$ denotes an orthogonal projection and at the same time its range.  $[A,B]$
is the commutator $AB-BA$ of the operators $A$ and $B$. The adjoint of $A$ is 
denoted as $\herm{A}$.  Furthermore, we shall use regularly the following well known facts.
\begin {itemize}
	\item Let $\Gamma$ be a positively oriented loop in the complex plane
 		circling the spectrum
		associated with an orthogonal eigenprojection $P$ of a self-adjoint
		operator $H$. Then $P=-(2\pi i)^{-1} \oint_\Gamma (H-z)^{-1}dz$. 
	
	\item Let $P(s)$ be a smooth family of orthogonal projections and
  		$Q(s)=1-P(s)$ the projection on the orthogonal
  		complement. Differentiating the relation $P^2(s)=P(s)$ gives  
		\begin{equation} \label{projoffdiag}
 			\dot P(s)= P(s)\dot P(s) Q (s)+Q(s)\dot P(s) P(s).
		\end{equation}
 		An operator of this type is called
 		``off-diagonal.''
\end{itemize} 
A summary of notation is given at the end of the paper (Table~\ref{tabnot}). 

Let $I \subset \RR$ be an interval with $0 \in I$  and $H(s), s \in I,$ 
a family of Hamiltonians.  We are interested in the time-dependent 
Schr\"odinger equations~(\ref{unscaled}),  
where $\tau>0$ defines a time scale. A change of variables converts
the family of equations~(\ref{unscaled}) to the equivalent family~(\ref{schroed}). 
We will denote the unitary time evolution associated
with Eq.~(\ref{schroed}) by $U_\tau(s)$. 

Throughout the article, $H$ is assumed to fulfill the following conditions.
\begin{enumerate}[(i)]
	\item $H(s)$, $s\in I$,  are self-adjoint operators on a Hilbert space
	   $\Hi$, with an $s$-independent domain $D$.
	\item$H(s)$ is a $k\geq 2$ times continuously differentiable map from 
	   $I$ to the space $ B(D,\Hi)$  of bounded linear operators from $D$ to $\Hi$ 
	   equipped with the graph norm of $H(0)$.
	\item $H(s)$ has gaps in the spectrum, and $P(s)$ is the
	   spectral projection on a band bordered by gaps, i.e., there are two real-valued,
   	   continuous functions $b_+$ and $b_-$,  and a number $g>0$ such that 
	$$ \dist \Bigl(\{b_+(s),\ b_-(s)\}, \sigma\bigl(H(s)\bigr) \Bigr)>g  \quad (s\in I),$$
	   and $P(s)$ is associated with the nonempty band $\sigma(H(s))\cap[b_-(s), b_+(s)]$, 
	 $s\in I$.
\end{enumerate}
Conditions (i) and (ii) imply that the resolvent $s \mapsto (H(s)-z)^{-1}$
as a map from $I$ to $B(\Hi)$ is $k$ times continuously differentiable and
ensure the existence and uniqueness of the propagator $U_\tau(s)$ if $k\geq 1$
(see, e.g., Ref.~\onlinecite{rs}): $U_\tau(s)$ is unitary, strongly continuous in $s$,
maps $D$ onto $D$, $U_\tau(0)={\rm id}_\Hi$, and for each $\psi \in D, \,
\psi_\tau (s):= U_\tau(s)\psi \in C^1(I, \Hi)$ and satisfies Eq.~(\ref{schroed}). 
Condition (iii) is illustrated by Fig.~1.

The first two adiabatic theorems give an estimate of the probability of finding
a state $\psi$ initially in $P(0)$ outside $P(s)$ at time $s$  and
compare $P_\tau(s):=U_\tau(s)P(0)\herm{U_\tau(s)}$ to $P(s)$ for 
large time scales $\tau$. $P_\tau(s)$ is the time-evolved 
projection, whereas $P(s)$ is the instantaneous spectral projection
associated with the band of the spectrum of interest. 

Let us remark that although bounds on the transition probability follow
directly from bounds on $||P_\tau(s)-P(s)||$, as can be seen
from relations~(\ref{transition}) and~(\ref{comparison}) below,
they are of sufficient importance that we state them separately in the theorems 
that follow.

\begin{theorem} [First adiabatic theorem]\label{adiab1}
	Assuming that the general  conditions on $H$ stated above are fulfilled, the following holds. 
	For $\psi \in P(0)$ the transition probability is of order  $O(1/ \tau^2)$, i.e.,
	\begin{equation*}
		\bigr(U_\tau(s)\psi, [1-P(s)]U_\tau(s)\psi \bigl)= O(1/ \tau^2)
	\end{equation*}
	and
	\begin{equation*}
                  ||P_\tau(s)-P(s)||=O(1/ \tau).
	\end{equation*}
	Both estimates hold uniformly on compact subsets of $I$. 
\end{theorem}

\begin{theorem} [Second adiabatic theorem\tp/\tp switching theorem] \label{adiab2}
	If in addition to the general conditions we assume  
	that $\dot H$ is compactly supported in an interval $]0,1[\subset I$,
	the above estimates can be improved to 
 	\begin{align*}
    		\bigl(U_\tau(s)\psi,[1-P(s)]U_\tau(s)\psi \bigr)&=O(1/\tau^{2(k-1)}),\\
      		||\thinspace P_\tau(s)-P(s)\thinspace ||&=O(1/ \tau^{k-1}),
 	\end{align*}
 	uniformly in $I\backslash \, ]0,1[$. 
\end{theorem}

\noindent{\emph{Remarks concerning Theorem~\ref{adiab2}:}}
	If $\dot H$ is compactly supported in an interval $]\alpha,\omega[\subset I$,
	the previous estimates change into the following:
	For $\psi \in P(\alpha)$,
	\begin{align*}
		\bigl(U_\tau(s;\alpha)\psi,[1-P(s)]U_\tau(s;\alpha)\psi \bigr)
			&=O({1/\tau^{2(k-1)}}),\\ 
 		||U_\tau(s;\alpha) P_\tau(s) \herm{U_\tau(s;\alpha)} -P(s)||
			&=O({1/ \tau^{k-1}}),
	\end{align*}
	uniformly in $I\backslash \, ]\alpha,\omega[$. The unitary $U_\tau(s;\alpha)$ denotes the
	propagator of the Schr\"odinger equation for the inital value $\alpha$.  

	For even more regular Hamiltonians, different methods of proof can be used to show that 
	the transition amplitude in Theorem~\ref{adiab2} becomes exponentially small in $\tau$. 
	For example, when the family of Hamiltonians $H(s)$ is in a Gevrey class, 
	one can apply methods of phase space tunneling \cite{j}, or use Nenciu's technique \cite{n2},
	exponentially small bounds are also given in Refs.~\onlinecite{hj} and~\onlinecite{jp}.

\begin{theorem} \label{theorem-gap1}
	Suppose that the spectrum of $H(s)$ restricted to $P(s)$ consists of $m(s)$
	eigenvalues (each possibly degenerate, crossing permitted) separated by a gap $g(s)$ from
	the rest of the spectrum of $H(s)$,  and $H$, $\dot H$, and $\ddot H$ are bounded
	operators. (This assumption is always fulfilled in finite-dimensional
	spaces.)  Then for $\psi\in P(0)$ and any $s\in I$,
	\begin{equation}\label{auxquantity}
		\bigl|\bigl(U_\tau(s)\psi,\thinspace [1-P(s)]U_\tau(s)\psi\bigr)\bigr|\leq A(s)^2, \quad 
		||P_\tau(s)-P(s)||\leq A(s),
	\end{equation}
	where 
	\begin{align*}
		A(s)&= {1\over \tau}\ub{{\sqrt{m} ||\dot P||\over
  		  	g}}  + {1\over \tau} \int_0^s (
  		  	\sqrt{m}{||Q\ddot PP||+||\dot 
			P||^2\over g }  + 2 m {||\dot H|| \thinspace ||\dot P||\over  g^2})\\
 		    & \leq {1\over \tau} \ub{{m ||\dot H||\over
  			g^2}}
 			+{1\over \tau} \int_0^s ({m||\ddot H||\over g^2} + 7 m\sqrt{m} {||\dot
   			H||^2\over g^3}), 
	\end{align*}
	and $f|_\mathrm{u.b.}$ is a shorthand for $f(0)+f(s)$. 
\end{theorem}

\begin{theorem}\label{theorem-gap2}
 	Suppose that the assumptions of Theorem~\ref{theorem-gap2} are satisfied and $H$ is 
	three times differentiable. Let $h(s):=\max(||\dot H(s)||, ||\ddot H(s)||, ||
	(d^3/ds^3) H(s)||)$. Then there is a constant $C$ (possibly
	depending on $m$) such that Eq.~(\ref{auxquantity}) holds with 
	\begin{equation*}
 		A(s)= \ub{{mh\over \tau g^2}} +{C\over \tau^2}\, \biggl(\ub{{h^2\over
   			g^4}} +{h\over g^2}(0)\int_0^s {h^2\over g^3} +\int_0^s{h^2\over g^5}
		+\int_0^s ds' {h^2\over g^3}(s')\int_0^{s'}ds'' {h^2\over g^3}(s'')\biggr).
	\end{equation*}
\end{theorem}

\section{Notation and identities}

\subsection{Adiabatic time evolution and wave operator}\label{ATE}

The proof of the adiabatic theorems is most easily accomplished with the aid of
an idealized time evolution, the adiabatic time evolution introduced in
Ref.~\onlinecite{asy1}, mapping  the spectral subspace $P(0)$ onto $P(s)$. This is an 
adaptation of an idea by Kato to the present situation \cite{kato}.

From the gap condition (iii) and $(H(\cdot)-z)^{-1} \in C^k(I,B(\Hi))$, it
follows that $P(s)$ is in $C^k(I,B(\Hi))$ as well. 
Define the adiabatic Hamiltonian associated with $P$ and the time
scale $\tau$ by
$$H_\tau^A(s):= H(s)+{i\over \tau}[\dot P(s), P(s)],\quad s\in I.$$
$H_\tau^A$ satisfies the conditions (i) and (ii) with
differentiability of order $k-1$. Existence of the
corresponding time evolution $U_\tau^A(\cdot)$ is ensured for $k-1 \geq 1$.

The adiabatic time evolution generated by $\tau H_\tau^A(s)$ is 
ideal in the following sense.
\begin{lemma} 
	$U_\tau^A(\cdot)$ satisfies the intertwining
  	property
	\begin{equation}\label{twin}
 		U_\tau^A(s) P(0)=P(s)U_\tau^A(s),\quad s\in I.
	\end{equation}
\end{lemma}

\begin{proof}[Sketch of proof] The strategy of the proof is to show that both
	sides satisfy the same initial value problem $X(0)=P(0)$, $\dot
	X(s)=-i\tau H_\tau^A(s)X(s)$ using that $\dot P$ is off-diagonal,
	see identity~(\ref{projoffdiag}) (see also Ref.~\onlinecite{asy1}, p. 38).
\end{proof}

In the following, we will omit the $s$-dependence and use the notation $Q=1-P$,
$P_0$, $Q_0$ as a shorthand for $Q(s) = 1-P(s)$, $P(0)$ and $Q(0)$. 

We prove the theorems by comparing the real time 
evolution $U_\tau(s)$ with the idealized time evolution
$U_\tau^A(s)$. 
The wave operator $\Om_\tau(s):=\herm{U_\tau^A(s)}U_\tau(s)$
 is a standard tool used in scattering theory to compare
dynamics, here, $U_\tau$ and $U_\tau^A$. 
We expect that adiabatic and
real time evolutions are close, i.e., $\Om_\tau$ is close to the identity,
for big $\tau$.  The proof
of the adiabatic theorems reduces to the estimation of the off-diagonal
blocks of the wave operator since for
$\psi \in P_0,\ ||\psi||=1$, 
\begin{equation}\label{transition}
(U_\tau\psi, QU_\tau\psi)=|| Q U_\tau P_0 \psi||^2
       =||QU_\tau^A\Om_\tau P_0\psi||^2\\
   \leq  ||Q_0\Om_\tau P_0||^2 
\end{equation}
and
\begin{equation} \label{comparison}
  	||P_\tau-P||=||U_\tau P_0 \herm{U_\tau}-U_\tau ^AP_0\herm{(U_\tau ^A)}||
   		=||\Om_\tau P_0-P_0\Om_\tau||
   		=||Q_0\Om_\tau P_0-P_0\Om_\tau Q_0||.
\end{equation}
By a straightforward calculation $\Om_\tau$ satisfies the Volterra
integral equation
\begin{equation} \label{volterra} 
 	\Om_\tau(s)=1-\int_0^s K_\tau(s')\Om_\tau(s')ds',
\end{equation}
where the kernel $K_\tau$ is defined by
$$K_\tau(s):=\herm{U_\tau^A(s)}[\dot P(s),P(s)]U_\tau^A(s),$$
and the integral here and in the following is the Riemann integral in the
strong sense.

\subsection{Integration by parts lemma}  \label{sect:IP}

With a map $X$ from $I$ to $B(\Hi)$, we associate another map $X[\cdot]$ defined
by $$X[s]:=\herm{U_\tau^{A}(s)} X(s)U_\tau^A(s).$$ With this notation the integral
kernel $K_\tau$ of the Volterra equation~(\ref{volterra}) introduced in the
last section is just  $[\dot P,P][\cdot]$ and the Volterra equation gives rise to integrals of 
the form $\int X[t]Y(t)dt$. Such integrals can be integrated by parts. \\

In the integral by parts formula (Lemma~\ref{ibp}), the twiddle operation plays
a crucial role. It can be understood as a partial inverse of the add
operation $A \mapsto [H,A]$ as follows.
\begin{lemma}  [Twiddle operation] \label{lemma-twiddle} Let $X$ be a
 	 map from  $I$ to 
  	$B(\Hi)$ and $\Gamma(s)$ a contour in $\mathbb{C}$ such that
  	$P(s)= - (2\pi i)^{-1}\oint_{\Gamma(s)} [H(s)-z]^{-1}dz$. Define 
	$$\tilde X(s): = {1\over 2\pi i}\oint_{\Gamma(s)}
		[H(s)-z]^{-1}X(s)[H(s)-z]^{-1}dz.$$ 
	Then $P\tilde X P=Q\tilde X Q=0$ and on $D$ 
	\begin{gather*}
 		[H, \tilde X]=PX-XP \\
 		P[H_\tau^A,\tilde X]Q=PXQ,\quad Q[H_\tau^A,\tilde
    		X]P=-QXP.
	\end{gather*} 
\end{lemma}
 
\begin{proof} We drop the $s$-dependence.
	Let $\Gamma'$ be a contour in $\mathbb{C}$ lying outside 
	the domain delimited by $\Gamma$ and such that $P$ and $\tilde X$ can
	be written as integrals 
	over $\Gamma'$ instead of $\Gamma$.  Then
	\begin{align*}
		\tilde X Q &= -{1\over (2\pi i)^2}\oint_\Gamma (H-z)^{-1}  X (H-z)^{-1}\tp  
        		    \oint_{\Gamma'}[{1\over z-z'}-(H-z')^{-1}]dz' dz\\
  			&=-{1\over (2\pi i)^2}\oint_\Gamma (H-z)^{-1} X (H-z)^{-1}\tp  
   			     \oint_{\Gamma'}{1\over z-z'}(H-z)^{-1}(H-z')^{-1}dz' dz\\
  			&=-{1\over (2\pi i)^2}\oint_{\Gamma'} \oint_\Gamma{1\over z-z'}(H-z)^{-1}
  			   dz X (H-z')^{-1}dz'\\
  			&= {1\over 2\pi i} \oint_{\Gamma'} P(H-z')^{-1}X(H-z')^{-1} dz'=P\tilde X. 
	\end{align*}
	This implies $P\tilde XP=0=Q\tilde X Q$. The commutator $[H,\tilde X]$
	can be computed on  
 	$D \supset\Ra(\tilde X)$,
	\begin{align*} 
		[H,\tilde X]&={1\over 2\pi i} \oint_{\Gamma} [H-z,(H-z)^{-1} X (H-z)^{-1}]dz\\
 			&={1\over 2\pi i}\oint_\Gamma \{X(H-z)^{-1}-(H-z)^{-1}X\}dz = -XP+PX.
	\end{align*}
	The second pair of identities follows with 
	$$P[[\dot P,P],\tilde X]Q=(P[\dot P,P]P)\tilde XQ-P\tilde X (Q[\dot P,P]Q)=0,$$
	and similarly $Q[[\dot P,P],\tilde X]P=0$. 
\end{proof}

\begin{lemma} [Partial integration] \label{ibp}
	Let $X\in C^1(I,B(\Hi))$, $Y\in C^0(I,B(\Hi))$.  Then $\tilde X$
	defined as above  
	is in $C^1(I,B(\Hi))$ and for $s \in I$, 
	\begin{multline*}
		\int_0^s P_0 X[s']Q_0 Y(s')ds' 
			=-{i\over \tau}\biggl(P_0\tilde
			X[s']Q_0Y(s')\Bigr|_0^s 
			-\int_0^s P_0\dot{\tilde X}[s']Q_0 Y(s')ds'\\
		-\int_0^s P_0\tilde 
			X[s']Q_0 \dot Y(s')ds'\biggr).
	\end{multline*}
	A similar equality holds with $P_0$ and $Q_0$
	interchanged and an overall change of sign on the right-hand side. 
 \end{lemma}

\noindent \emph{Remark.} The notation $X[s]$ is a shorthand for adjoining
$\herm{(U_\tau^A)}$ and $U_\tau^A$ \emph{after} taking into account other
symbols on $X$, e.g. 
$\dot {\tilde X}[s]= \herm{U_\tau^A(s)} \dot{\tilde X}(s) U_\tau ^A(s)$. 

\begin{proof}
	It suffices to notice that $\tilde X[\cdot]$ is strongly
	differentiable and  
	\begin{align*}
		{d\over ds}P_0 \tilde X[s] Q_0
		&=P_0 (i\tau [H_\tau^A,\tilde
			X][s]+\dot{\tilde X}[s])Q_0 
			=(i\tau P[H_\tau^A,\tilde X]Q)[s]+P_0\dot{\tilde X}[s]Q_0 \\
	   	&=P_0 (i \tau X[s]+\dot{\tilde X}[s])Q_0,
	\end{align*}
	where the intertwining property has been used in the form
	\begin{equation*}
	 (PZQ)[s]=P[s]Z[s]Q[s]= P_0 Z[s] Q_0
	\end{equation*}
	 for $Z:\ I\rightarrow B(\Hi)$.
\end{proof}

\section{Proof of Theorem~\ref{adiab1} and Theorem~\ref{adiab2}}

Using the intertwining property, we see that the integral kernel of
the Volterra equation~(\ref{volterra}) can be written as
\begin{equation*}
 	K_\tau(s)=(\dot P P- P\dot P)[s] = (Q \dot P P -P\dot P Q)[s]
  		= Q_0 \dot P[s] P_0 - P_0 \dot P[s] Q_0.
\end{equation*} 
As a consequence, the Volterra equation~(\ref{volterra}) written for
$Q_0\Om_\tau P_0$ becomes
\begin{equation*}
	Q_0\Om_\tau(s) P_0 = -\int_0^s Q_0 \dot P[s'] P_0 \Om_\tau(s')P_0ds'.
\end{equation*}
We can apply Lemma~\ref{ibp} and integrate by parts 
\begin{align}\label{ibp-1}
	Q_0\Om_\tau(s) P_0&=-{i\over \tau}\Bigl( Q_0 \tilde{\dot P}[s']P_0
  		\Om_\tau(s')P_0|_0^s - \int_0^s Q_0 \dot{\tilde {\dot P}}[s']P_0
  		\Om_\tau(s') P_0 ds' \notag \\
 	&\quad - \int_0^s Q_0 \tilde{\dot P}[s'] P_0 \dot
  P[s'] Q_0 \Om_\tau(s')P_0 ds'\Bigr).
\end{align}
Therefore $||Q_0\Om_\tau P_0||=O(1/\tau)$. The same method gives the
bound $||P_0\Om_\tau Q_0||=O(1/\tau)$, so that Theorem~\ref{adiab1}
follows with Eqs.~(\ref{transition}) and~(\ref{comparison}). 

Now, suppose that $\dot H$ is compactly supported in $]0,1[$. The Volterra
equation~(\ref{volterra}) shows $\Om_\tau(s)=1$ for $s\leq 0$ and
$\Om_\tau(s)=\Om_\tau(1)$ for $s\geq 1$; hence, it is enough to prove
the statement for $s=1$. 
In Eq.~(\ref{ibp-1}), for $s\geq 1$, the boundary terms vanish.  
The basic idea to prove Theorem~\ref{adiab2} is to estimate the two
remaining integrals by iterating the 
integration by parts. However, doing so naively for the second integral
with $X(s)=\tilde{\dot P}(s)$ and $Y(s)=P[s]Q_0 \Om_\tau(s)P_0$ 
leads to derivatives of $P[s]=\herm{U_\tau^A(s)} P(s)U_\tau^A(s)$ and thus
to powers of $\tau$. This can be avoided by inserting the expression
for $Q_0\Om_\tau P_0$ recursively into the second integral. As a price
to pay, we have to consider multiple integrals. \label{problem}

\begin{lemma} \label{cjl}
	Let $0\leq m\leq k$. Then for all $j\in \mathbb{N}$ and $X_1,...,X_j\in
		C_0^m\bigl(]0,1[,B(\Hi)\bigr)$ the integral 
 	$$A_j(X_1,...,X_j)=Q_0\int_0^1 ds_1X_1[s_1]Q_0..
		.. Q_0\int_0^{s_{j-1}} ds_j X_{j}[s_j]P_0
  	\Om_\tau(s_j) P_0 $$
	is of order $O(1/\tau^m)$.
\end{lemma}

\noindent \emph{Remark.} In this notation, $Q_0 \Om_\tau(1)P_0=-A_1(\dot P)$. 

\begin{proof} By induction over $m$. The statement for $m=0$ is
	obvious since all $A_j$ are bounded as functions of $\tau$. For the
	induction step $m\leq k-1 \rightarrow m+1$ assume that $X_1,..,X_j$ are in
	$C_0^{m+1}(]0,1[,B(\Hi))$. Suppose first that $j\geq 2$. We integrate by
	parts   with respect to the variable $s_j$ and use the vanishing of
	boundary terms,
	\begin{align*}
		&\int_0^{s_{j-1}} Q_0 X_j[s_j]P_0 \Om_\tau(s_j)P_0ds_j\\
		&\quad ={i\over \tau} \biggl(Q_0\tilde X_j[s_{j-1}] P_0 \Om_\tau(s_{j-1})P_0 
 		    - \int_0^{s_{j-1}}Q_0 \dot{\tilde X}_j[s_j]P_0 \Om_\tau(s_j)P_0 ds_j\\
  		&\quad \quad - \int_0^{s_{j-1}}Q_0 \tilde X_j[s_j] P_0 \dot P[s_j]Q_0
 		    \Om_\tau(s_j)P_0 ds_j\biggr)\\ 
		&\quad  = {i\over \tau} \biggl( Q_0\tilde X_j[s_{j-1}] P_0 \Om_\tau(s_{j-1})P_0 
 		     - \int_0^{s_{j-1}}Q_0 \dot{\tilde X}_j[s_j]P_0 \Om_\tau(s_j)P_0 ds_j\\
 		& \quad \quad  + \int_0^{s_{j-1}} ds_j Q_0 (\tilde X_j P \dot P)[s_j]Q_0 
 			\int_0^{s_j} ds_{j+1} \dot P[s_{j+1}]  P_0 \Om_\tau(s_{j+1}) P_0\biggr).
	\end{align*}
	Inserting this into the definition of $A_j$, we get
	\begin{align*}
 		A_j(X_1,...,X_j)
		&={i\over \tau}\biggl(A_{j-1}(X_1,...,X_{j-2},X_{j-1} Q \tilde
 			X_j)- A_j(X_1,...,X_{j-1}, \dot{\tilde X}_j )\\
		&\qquad \qquad - A_{j+1} 
 		  (X_1,...,X_{j-1}, \tilde X_j P \dot P, \dot P)\biggr). 
	\end{align*}
	This remains valid for $j=1$ if we set $A_0:=0$. 
	The new variables $X_{j-1}Q \tilde X_j$, $\dot{\tilde
  		X}_j$, $\tilde X_j P \dot P$, and $\dot P$ 
	are in $C^m_0\bigl(]0,1[,B(\Hi)\bigr)$ (here we use $m\leq k-1$). 
 	Hence,
	\begin{equation*}
	A_j(X_1,..,X_j)={i\over
  		\tau}\,O\left({1\over\tau^m}\right)=O\left({1\over \tau^{m+1}}\right). \hfill \qedhere
	\end{equation*}
\end{proof}

\begin{proof}[Proof of Theorem~\ref{adiab2}]
	Suppose that $\dot H$ is compactly supported in $]0,1[$. Then $\dot P$
	belongs to $C_0^{k-1}(]0,1[,B(\Hi))$. Lemma~\ref{cjl} implies 
	$Q_0\Om_\tau(1)P_0 =-A_1(\dot P)=O(1/\tau^{k-1})$. The lemma still
	holds with $Q$ and $P$ interchanged so
	$P_0\Om_\tau(1)Q_0=O(1/\tau^{k-1})$. Together with Eqs.~(\ref{transition})
	and~(\ref{comparison}), this concludes the proof. 
\end{proof}

\section{Dependence on the gap}

Up to now, we have examined the dependence of the error terms on the time scale
$\tau$ only. In order to give error bounds with explicit gap dependence, we
need to know how $P$ decorated with dots and tildes behaves. The aim of the
following series of lemmas is to generalize the bounds given in Ref.~\onlinecite{reich},
pp.~8--9, to the case where $P$ projects on more than one eigenvalue. These
allow to prove Theorems~\ref{theorem-gap1} and~\ref{theorem-gap2}. The reader
interested in projections $P(s)$ associated with a single (possibly degenerate)
eigenvalue can find a simpler proof of the inequalities stated in
Lemma~\ref{lemma-gap4}, up to a constant factor, in Ref.~\onlinecite{reich}. 

We will assume throughout this section that all operators are bounded and that
$PHP$ has discrete spectrum. This is always fulfilled in finite-dimensional
spaces.  We write 
$$P(s)=\sum_{j=1}^{m(s)} P_j(s),\qquad  H(s)P_j(s)=\lambda_j(s) P_j(s), $$
where $P_j(s)$ are the projections associated with the $m(s)$ 
eigenvalues $\lambda_1(s)$, ... ,$\lambda_{m(s)} (s)$. Note that each
eigenvalue might be degenerate and the eigenvalues are allowed to cross. 
Furthermore, let 
$$g(s):=\dist\Bigl(\{
\lambda_1(s),...,\lambda_{m(s)}\},\ 
\sigma\bigl(H(s)\bigr)\backslash\{\lambda_1(s),..,\lambda_{m(s)} (s) \}\Bigr)$$
denote the gap in $s$. In the following, we will drop the
$s$ dependence and use the notation 
$$R_z=[H(s)-z]^{-1}, \qquad \hat R_z = Q \{[H(s)-z]\! \upharpoonright_Q \}^{-1} Q.$$
Note that the reduced resolvent $\hat R_z$ is well defined even if $z$
is one of the eigenvalues $\lambda_j$ and $||\hat R_{\lambda_j}||\leq
1/g$.

\begin{lemma} \label{lemma-gap1}
	Let $X,A,B$ be bounded operators. Then 
	\begin{gather}
		\tilde X=-\sum_{j=1}^m ( P_j X \hat R_{\lambda_j} + \hat R_{\lambda_j} X
			P_j)\label {tilde-X},\\ 
		{1\over 2\pi i} \oint_\Gamma R_z A R_z B R_z dz =(P-Q)( \tilde A
			\tilde B + \widetilde{A\tilde B} - \widetilde {\tilde A B}) \label{G(A,B)}.
	\end{gather}
\end{lemma}

\begin{proof} By Lemma~\ref{lemma-twiddle}, $[H,X]=[P,X]$. This
	implies 
	$$P_j [H,X] Q = \lambda_j P_j XQ- P_j XQHQ = P_jXQ.$$
	Hence,
	$$P XQ= \sum_j P_j XQ = -\sum_j P_jX \hat R_{\lambda_j}.$$
	Similarly,
	$QXP =-\sum _j \hat R_{\lambda_j} X P_j$. 
 	Since by Lemma~\ref{lemma-twiddle} $\tilde
	X$ is off-diagonal, Eq.~(\ref{tilde-X}) follows. 
	Now, let
	\begin{equation*}
 		G(A,B):= {1\over 2\pi i} \oint_\Gamma R_z A R_z BR_z dz.
	\end{equation*}
	Then 
	\begin{equation*} 
		[H,G(A,B)]={1\over 2\pi i}\oint_\Gamma [H-z, R_z A R_z B R_z] dz 
 		= A \tilde B - \tilde A B.
	\end{equation*}
	An argument similar to the  reasoning above fixes the off-diagonal part of
	$G(A,B)$, 
	\begin{align}\label{PGQ}
		PG(A,B)Q&=P (\widetilde{A\tilde B}-\widetilde{\tilde AB})Q,
				\qquad 
			QG(A,B)P=-Q (\widetilde{A\tilde B}-\widetilde{\tilde AB})P.
	\end{align}
	The diagonal blocks of $G(A,B)$ can be computed, using that 
	$z\mapsto \hat R_z$ is holomorphic in the domain enclosed by $\Gamma$ and
	has  $\hat R_z^2$ as its derivative with respect to $z$,
	\begin{align}
		QG(A,B)Q&=\sum_j {1\over 2\pi i} \oint_\Gamma {1\over \lambda_j -z}
			\hat R_z  \notag
			A P_j B \hat R_z dz+ {1\over 2\pi i} \oint_\Gamma \hat R_z A\hat R_z B
			\hat R_z dz \\
 		     &= -\sum_j \hat R_{\lambda_j} A P_j B \hat R_{\lambda_j} +0 
 			= - Q\tilde A \tilde B Q.\label{QGQ}
	\end{align}
 	Similarly, 
	\begin{align*}
		PG(A,B)P&=\sum_{j,k} {1\over 2\pi i} \oint_\Gamma {1\over \lambda_j
  			 	-z}{1\over \lambda_k-z} P_j A \hat R_z B P_k dz \\
  			&\qquad +\sum_{j,k,l}{1\over 2\pi i}\oint_\Gamma
  				{1\over (\lambda_j-z)(\lambda_k-z)(\lambda_l-z)}dz\ P_j A
  				P_k B P_l .
	\end{align*}
	Now, 
	$${1\over 2\pi i} \oint_\Gamma {1\over( \lambda_j-z)^2} \hat R_z =
		R_{\lambda_j}^2.$$ 
 	For $j\neq k$, 
	\begin{align*} 
		{1\over 2\pi i}  \oint_\Gamma {1\over (\lambda_j-z)(\lambda_k-z)}R_z
		&={1\over \lambda_j -\lambda_k} (\hat R_{\lambda_j}-\hat R_{\lambda_k})
 			= \hat R_{\lambda_j}\hat R_{\lambda_k},
	\end{align*}
	and $\oint_\Gamma [(\lambda_j-z)(\lambda_k-z)(\lambda_l-z)]^{-1} dz=0$
	for any $j, k,l\in \{1,...,m(s)\}$, 
	so that 
	\begin{equation} \label{PGP}
		PG(A,B)P= \sum_{j,k} P_j A \hat R_{\lambda_j} \hat R_{\lambda_k} B
		P_k =  P \tilde A \tilde B P.
	\end{equation}
	Putting  together Eqs.~(\ref{PGQ})--(\ref{PGP}) and making
	use of the off-diagonal character of any ``twiddled'' operator $\tilde
	A$, $\tilde B$, etc., gives Eq.~(\ref{G(A,B)}). 
\end{proof}

\begin{lemma} \label{lemma-gap2}
	We have $\dot P=\tilde{\dot H}$. 
	Let $X$ be a continuously differentiable map from $I$ to
  	$B(\Hi)$. Then
	\begin{equation} \label{DGamma}
		\dot{\tilde X}=\tilde {\dot X}+(Q-P)(\dot P \tilde X + \tilde X \dot
		P + \widetilde{[\dot H, \tilde X]}-\widetilde{[\dot P,X]}).
	\end{equation}
	In particular,  
	\begin{equation} \label{DGammaDP}
		Q\dot{\tilde{\dot P}}P= Q(\tilde{\ddot P} + \widetilde{\dot
  		H\tilde{\dot P}}-\widetilde{\tilde{\dot P}\dot H})P.
	\end{equation}
\end{lemma}

\begin{proof} First notice 
	$$\dot P =-{d\over ds}{1\over 2\pi i}\oint_\Gamma R_z dz = {1\over
  		2\pi i}  \oint R_z
		\dot H R_z dz = \tilde{\dot H}.$$
	Identity~(\ref{DGamma}) then follows with 
	\begin{align*}
		\dot{\tilde X}&= {d\over ds}{1\over 2\pi i} \oint_\Gamma  R_z X R_z dz
 			= {1\over 2\pi i} \oint_\Gamma (R_z \dot X R_z - R_z \dot H R_z X R_z
 		- R_z X R_z \dot H R_z ) dz
	\end{align*} 
	and Lemma~\ref{lemma-gap1}. Since $\dot P \tilde X$ and $\tilde X \dot
	P$ are diagonal operators and $[\dot P,\dot P]=0$,  Eq.~(\ref{DGammaDP})
	follows from Eq.~(\ref{DGamma}) applied to $X=\dot P$. 
\end{proof}

\begin{lemma} \label{lemma-gap3}
	Let $X:[0,1]\rightarrow B(\Hi).$ Then
	\begin{equation} \label{normtwiddle}
 		||\tilde X(s)||\leq \sqrt{m(s)}{||X(s)||\over g(s)}. 
	\end{equation}
\end{lemma}

\begin{proof} Let $\phi\in \Hi$. 
	\begin{align*}
		||\tilde X\phi ||=||(P\tilde X Q +Q\tilde XP)\phi||^2 
		&= ||P\tilde X Q\phi||^2 + ||Q\tilde XP \phi||^2 \\
		& \leq \max(||P\tilde XQ||^2, ||Q\tilde X P||^2)||\phi||^2.
	\end{align*}
	Moreover, 
	$$Q\tilde X P =-\sum_{j=1}^m \hat R_{\lambda_j} X P_j = \herm{(P\tilde{\herm{X}} Q)}$$ 
	and for bounded operators $||A||=||\herm{A}||$, hence it is enough to prove
	$$||PXQ||\leq \sqrt{m} {||X||\over g}.$$
	But this follows from $||PXQ||=||\herm{(PXQ)} (PXQ)||^{1/2}$  and 
	\begin{align*}
		||PXQ||^2&=||\herm{(PXQ)} (PXQ)|| = ||\sum_{j=1}^m \hat R_{\lambda_j} \herm{X}P_j X
				\hat R_{\lambda_j}||\\
 			&\leq m {1\over g^2} ||X||\thinspace ||\herm{X}|| = m {||X||^2\over g^2}.
	\qedhere
	\end{align*}
\end{proof}

\begin{lemma} \label{lemma-gap4}
	The following series of bounds holds: 
	\begin{align*}
 		||\dot P||&\leq \sqrt{m} {||\dot H||\over g}\\
 		||\tilde {\dot P}||&\leq \sqrt{m} {||\dot P||\over g}\leq m {||\dot
   		H||\over   g^2}\\
 		||Q\ddot P P ||&\leq \sqrt{m} {||\ddot H||\over g}+{4m ||\dot
   		H||^2\over g^2}\\ 
 		||Q\dot {\tilde {\dot P}}P||&\leq \sqrt{m} {||Q\ddot P P||\over g}+{2
   			m ||\dot 
   		H||\thinspace ||\dot P||\over g^2}  
  		\leq {m ||\ddot H||\over g^2}+6{m \sqrt{m} ||\dot H||^2\over g^3}. 
	\end{align*}
	More generally, let $\Gamma X:=\tilde X$, $DX:= \dot X$, then 
	if $H$ is $(l+1)$-times differentiable and 
	$h(s):= \max_{1\leq j\leq l+1} ||({d\over ds})^jH(s)||$, there 
 	exists a constant $C_l$, possibly depending on the number $m$ of
 	eigenvalues, such that 
	\begin{align}  \label{normtwiddle-it}
		||(D\Gamma)^l D P||&\leq {C_l\over g^{l}}
		\sup_{\substack{k,\alpha_1,...\alpha_l \geq 1:\\ \alpha_1+..+\alpha_k=l+1}}\
		 \prod_{1\leq j\leq k} {1\over g} 
		||{d^{\alpha_j}  H \over ds^{\alpha_j}}||\leq {C_l\over g^l} ({h\over
  		g})^{l+1}, 
	\end{align} 
	i.e., every tilde pulls out a $1/g$ and every dot an $h/g$. 
\end{lemma}

 The proof of the series of bounds is straightforward
 with Lemma~\ref{lemma-gap2} and Eq.~(\ref{lemma-gap3}). The estimate
 (\ref{normtwiddle-it}) is shown with an induction over $l$. \\

Theorem~\ref{theorem-gap1} now is a simple consequence of
Eq.~(\ref{ibp-1}) and Lemma~\ref{lemma-gap3}. If $H$ is three times
differentiable, iterating the integration by parts gives the expansion

\begin{equation} \label{2d-order}
 \begin{split}
	&Q_0\Om_\tau (s) P_0 \\
	&\quad ={i\over \tau} Q_0 \pone [s'] \Om_\tau(s')
		P_0|_0^s + {1\over   \tau^2} 
 		Q_0 \ptwo [s']\Om_\tau(s') P_0 |_0^s  \\
	&\qquad -{1\over \tau^2}\int_0^s
 		Q_0 (\pone \dot P )[s'] ds') \pone [0] P_0
   		-{1\over \tau^2}\int_0^s Q_0(\dot {\ptwo}P+\ptwo \dot P -\pone 
 		\dot P \pone )[s']\Om_\tau(s') P_0ds' \\
	& \qquad +{1\over \tau^2} \int_0^s  Q_0 (\pone \dot P )[s']\int_0^{s'} 
     		(\dot{\pone }P+ \pone \dot P)[s''] \Om_\tau(s'') P_0 ds''ds'.
 \end{split}
\end{equation}
Theorem~\ref{theorem-gap2} follows from this expansion and
Lemma~\ref{lemma-gap4}.

\noindent \emph{Comment on the traditional ``adiabatic criterion.''} 
	How big must $\tau$ be in order to ensure
	that the error $||Q_0\Om_\tau P_0||$ remains smaller than some
	constant $\epsilon<1$? A frequently given answer \cite{m} is
	\begin{equation*}\label{adiabatic-criterion}
		\tau \geq {\rm const}\ \sup_{0\leq s\leq 1} {||\dot H(s)||\over
  		g(s)^2}.
	\end{equation*}
	This criterion comes from the first order term in the expansion~(\ref{2d-order}) 
	which we rewrite as
	$$||Q_0\Om_\tau(s)P_0||\leq{1\over \tau} \biggl( {m||\dot H||\over g^2}(0) +  
 	{m||\dot H||\over g^2}(s)\biggr) + {C(H) \over \tau^2}.$$
	However, $C(H)$ in  the second order term may depend on the gap. 
	For a fixed Hamiltonian, one can always justify neglecting higher order
	times by taking $\tau$ sufficiently large.     However,  in  applications to quantum computation
	one is often interested in a family of Hamiltonians in which the gap may change.
	Indeed,  in the situations considered in Theorems 3 and 4, the second order coefficient
	$C(H)$ can be $O(1/g^6)$. In such a case, this term is unbounded 
	for vanishing $g$ unless $\tau = O(1/g^3)$.

	There have been attempts toward a rigorous criterion based on an
	estimate of the form 
	\begin{equation} \label{reich_bound}
 		||Q_0\Om_\tau (s) P_0||\leq C\bigl(\ub{{h\over \tau g^2}} + \sup_{0\leq
		s'\leq s} {1\over \tau^k} {h^{k+1}\over g^{2k+1}}\bigr)
	\end{equation}
	for $(k+1)$-times differentiable Hamiltonians 
	(Ref.~\onlinecite{reich}, p.~2), where \newline 
	$\displaystyle{h(s)= \max_{0\leq l\leq k+1} 
	||\tfrac{d^lH}{ds^l} (s)||}$. This bound suggests that the transition amplitude can be kept 
	small by choosing $\tau$ of the order of $g^{2 +1/k}$. For large $k$, this is very 
	close to the traditional choice $\tau \sim g^2$ and much better than the scaling 
	$\tau\sim g^3$ obtained from Theorem~\ref{theorem-gap2}.

	The methods presented here do not allow to prove the bound~(\ref{reich_bound})
	 when $k\geq 2$ (compare to 
	Theorem~\ref{theorem-gap2}). Trying to prove it by iterating the integration by parts fails; as we saw in 
	Sec.~\ref{problem}, a direct iteration of the integration by parts generates powers of $\tau$ 
	(see p.~\pageref{problem}). This was overlooked in Ref.~\onlinecite{asy1} and later
	on corrected in Refs.~\onlinecite{asy2} and~\onlinecite{ks}. However, it seems that a bound of 
	type~(\ref{reich_bound}) can be proved using Nenciu's expansion, as sketched in Ref.~\onlinecite{elgart}.


\section{Application to interpolating Hamiltonians}

In adiabatic quantum computation, the Hamiltonians of interest are
interpolations between an initial Hamiltonian $H_0$ with an easily
computable ground state and a final Hamiltonian $H_1$ whose ground
state encodes the solution to some problem. The running time of an
``adiabatic algorithm'' is closely related to the time $\tau$ required
to ensure a good agreement between real and adiabatic time evolutions
and depends crucially on the minimal gap $g_{\mathrm{min}}$. 

At first glance, Theorem~\ref{theorem-gap1} does not imply 
the general wisdom that $\tau$ should be of order
$1/g_{\mathrm{min}}^2$: if we estimate the integrals by the maximum of the
integrand, we get a worst case dependence in $1/g_{\mathrm{min}}^3$. 
 A closer look at the following example however shows that it is not a good idea
to bound the integral in this way and that additional knowledge
about the gap function $g(s)$ \emph{does} allow to extract the $1/g_{\mathrm{min}}^2$ 
behavior from Theorem~\ref{theorem-gap1}. Moreover, as suggested  in
Refs.~\onlinecite{vmz,rc}, it is possible to improve the error dependence on the minimal
gap by adapting the interpolation between $H_0$ and $H_1$. \\

\emph{Example:} (See also Ref.~\onlinecite{vmz}) Let $|0\ra, |1\ra$ denote
an orthonormal basis of 
$\CC^2$. For \mbox{$w\in \{0,1\}^n$}, let $|w\ra = |w_1\ra \otimes\cdots\otimes
|w_n \ra$, $\Hi :=(\CC ^2)^{\otimes n}$,
$$H_0:=1-|\hat 0\rangle \langle \hat 0|, \quad H_1:=1-|u\rangle \langle 
u|,$$   
where $|\hat 0\ra = 2^{-n/2} \sum_{w\in \{0,1\}^n} |w\ra$ and $u\in
\{0,1\}^n$. The element $u$ is regarded as an unknown element to be searched for in a
list of length $2^n$. We claim the following.
\begin{enumerate}[(1)]
	\item Let $H(s)=(1-s)H_0+sH_1$. Then $g_{\mathrm{min}}=2^{-n/2}$, and there
  		is a constant $C>0$ such that 
  		$||Q_0\Om_\tau(1) P_0||\leq   C/(\tau g_{\mathrm{min}}^2)$.
	\item Let $H(s)=(1-f(s))H_0 + f(s) H_1$. It is possible to choose
  		$f:[0,1]\rightarrow [0,1]$ onto, 
  		monotone strictly increasing, such that 
 		$||Q_0\Om_\tau (1)P_0||\leq   C/(\tau g_{\mathrm{min}})$
 		for some constant $C>0$. 
\end{enumerate}

\begin{proof} $H(s)=(1-s)H_0+sH_1$ has a nondegenerate ground state
	separated from the first excited state by the gap 
	\begin{equation} \label{gap-example}
 		g(s) =\sqrt{2^{-n}+ 4(1-2^{-n})(s-\tfrac{1}{2})^2}.
	\end{equation} 
	It is $1$ at the boundaries $s=0,1$. We notice that for $p>1$, 
	$\int_0^1 g^{-p}(s) ds = O(g_{\mathrm{min}}^{1-p})$
	and  $||\dot H||=||H_1-H_0||\leq 2$. The first claim now follows from
	Theorem~\ref{theorem-gap1}. 

	Next, $H(s)=[1-f(s)]H_0 + f(s) H_1$ has gap $g\bigl(f(s)\bigr)$ with $g$ defined
	in Eq.~(\ref{gap-example}). Let $f:[0,1]\rightarrow [0,1]$ be the solution of
	$$f(0)=0, \quad \dot f(s) = kg^p\{f(s)\}, \quad k=\int_0^1
	g^{-p}(u)du.$$
	Then $f(1)=1$. If $p>1$, $k=O(g_{\mathrm{min}}^{1-p})$. Since 
	$$\dot H(s)=k g^p\bigl(f(s)\bigr) (H_1-H_0), \quad \ddot H(s)= k^2 g^{2p-1}\bigl(f(s)\bigr)
	\tp \dot g\bigl(f(s)\bigr)\tp (H_1-H_0),$$
	Theorem~\ref{theorem-gap1} gives 
	\begin{align*}
		||Q_0\Om_\tau(s)P_0|| & \leq {1\over \tau} \Bigl( 4 k + k^2 \int_0^1 \bigl[2
			g^{2p-3}\bigl(f(s)\bigr)\tp \bigl|\dot g\bigl(f(s)\bigr)\bigr|
		 + 28 g^{2p-3}\bigl(f(s)\bigr) \bigr]ds \Bigr) \\
 		& \leq {k\over \tau}\Bigl( 4  +  \int_0^1 \bigl[2 g^{p-3}(u)\tp |\dot g(u)|+ 
 			28 g^{p-3}(u) \bigr] du\Bigr) 
	\end{align*} 
	(we used the change of variables $u=f(s)$). For $p<2$, $\int
	g^{p-3}=O(g_{\mathrm{min}}^{p-2})$. The function $g$ is decreasing on $[0,1/2]$ and increasing
	on $[1/2,1]$; therefore for $p<2$
	$$\int_0^1 g^{p-3}|\dot g|= {2\over p-2}
	(1-g_{\mathrm{min}}^{p-2})=O(g_{\mathrm{min}}^{p-2}).$$
	Hence, for $1<p<2$, $||Q_0\Om_\tau(1)P_0||=O(\tau^{-1}g_{\mathrm{min}}^{1-p}
	g_{\mathrm{min}}^{p-2})=O((\tau g_{\mathrm{min}})^{-1})$. This concludes the proof. 
\end{proof}

More generally, the previous considerations concerning the gap
dependence will apply provided that the gap function $g(s)$ has the three
following properties: 
\begin{itemize}
	\item[-] The gap at the boundaries ($s=0, \ s=1$) is of order 1. 
	\item[-] The gap decreases strictly towards its minimal value
  		$g_{\mathrm{min}}$ and then increases strictly.  
	\item[-]  If $p>1$, then $\int_0^1 g^{-p}(s) ds =
  		O(g_{\mathrm{min}}^{1-p})$. 
\end{itemize} 
The first two features are shared by most examples studied so far
\cite{reich,sms}. The third one needs to be checked more
carefully, but 
it is reasonable to expect that $\int g^{-p}$ behaves better than
$g_{\mathrm{min}}^{-p}$. 

\begin{table}[here]
	\caption{ Summary of notation. \label{tabnot}}
  	$\begin{array}{ccl}
    \hline\hline
	\text{Symbol}&	& \text{Meaning} \\
	\hline	
	U_\tau(s)&  	&\textrm{real time evolution}\\
 	U_\tau^A(s)&  	&\textrm{adiabatic time evolution}\\
        X[s]& 	  	&\herm{U_\tau^A(s)}X(s) U_\tau^A(s) \\
	\tilde{X}(s)&  	& (2\pi i)^{-1}\oint_{\Gamma(s)} [H(s)-z]^{-1}X
        		[H(s)-z]^{-1} dz\\
	\Om_\tau(s)&  	&\herm{U_\tau^A(s)} U_\tau(s)\quad \textrm{(wave operator)}\\
	R_z& 		& [H(s)-z]^{-1}\quad \textrm{(resolvent)}\\
	\hat R_z& 	& Q \{[H(s)-z]\! \upharpoonright_Q \}^{-1} Q\quad \textrm{(reduced resolvent)}\\
    \hline\hline
   \end{array}$
\end{table}

\begin{acknowledgments}
	M.-B.~R. was partially supported  by
 	the National Security Agency (NSA) and
 	Advanced Research and Development Activity (ARDA) under
	Army Research Office (ARO) contract number 
     	DAAD19-02-1-0065, and by the National Science
        Foundation under Grant  DMS-0314228. R.~S. and S.~J. were 
	supported by the Deutsche Forschungsgemeinschaft (DFG).
\end{acknowledgments}


\end{document}